\documentclass[sigconf, nonacm]{acmart}

\usepackage{csquotes}
\usepackage{dirtytalk}

\AtBeginDocument{%
  \providecommand\BibTeX{{%
    \normalfont B\kern-0.5em{\scshape i\kern-0.25em b}\kern-0.8em\TeX}}}

\begin{document}

\title{Prompts First, Finally}

\author{Brent N. Reeves}
\orcid{0000-0001-5781-1136}
\affiliation{%
  \institution{Abilene Christian University}
  \city{Abilene}
  \state{TX}
  \country{USA}
}
\email{brent.reeves@acu.edu}

\author{James Prather}
\orcid{0000-0003-2807-6042}
\affiliation{
  \institution{Abilene Christian University}
  \city{Abilene}
  \state{TX}
  \country{USA}
}
\email{james.prather@acu.edu}

\author{Paul Denny}
\orcid{0000-0002-5150-9806}
\affiliation{
  \institution{The University of Auckland}
  \city{Auckland}
  \country{New Zealand}
}
\email{paul@cs.auckland.ac.nz}

\author{Juho Leinonen}
\orcid{0000-0001-6829-9449}
\affiliation{
  \institution{Aalto University}
  \city{Espoo}
  \country{Finland}
}
\email{juho.2.leinonen@aalto.fi}

\author{Stephen	MacNeil}
\affiliation{
  \institution{Temple University}
  \city{Philadelphia}
  \state{PA}
  \country{United States}}
\email{stephen.macneil@temple.edu}
\orcid{0000-0003-2781-6619}

\author{Brett A. Becker}
\orcid{0000-0003-1446-647X}
\affiliation{
  \institution{University College Dublin}
  \city{Dublin}
  \country{Ireland}
}
\email{brett.becker@ucd.ie}

\author{Andrew Luxton-Reilly}
\orcid{0000-0001-8269-2909}
\affiliation{
  \institution{The University of Auckland}
  \city{Auckland}
  \country{New Zealand}
}
\email{a.luxton-reilly@auckland.ac.nz}

\begin{abstract}

Generative AI (GenAI) and large language models in particular, are disrupting Computer Science Education. They are proving increasingly capable at more and more challenges. 
Some educators argue that they pose a serious threat to computing education, and that we should ban their use in the classroom.
While there are serious GenAI issues that remain unsolved, it may be useful in the present moment to step back and examine the overall trajectory of Computer Science writ large. 
Since the very beginning, our discipline has sought to increase the level of abstraction in each new representation. We have progressed from hardware dip switches, through special purpose languages and visual representations like flow charts, all the way now to ``natural language.'' 
With the advent of GenAI, students can finally change the abstraction level of a problem to the ``language'' they've been ``problem solving'' with all their lives. 
In this paper, we argue that our programming abstractions were always headed here -- to natural language. 
Now is the time to adopt a ``Prompts First'' approach to Computer Science Education.

\end{abstract}



\maketitle

%
%

\section{Introduction}

In 1975, while managing the development of OS/360 at IBM, Fred Brooks described programming\cite{brooks1975mythical}:

\begin{displayquote}
One must perform perfectly. The computer resembles the magic of legend in this respect, too. If one character, one pause, of the incantation is not strictly in proper form, the magic doesn't work. Human beings are not accustomed to being perfect, and few areas of human activity demand it. Adjusting to the requirement for perfection is, I think, the most difficult part of learning to program.
\end{displayquote}

In part to ease the pain of such punishing exactitude that Brooks observed, the history of Computer Science reveals an ongoing quest to raise the level of abstraction through the creation of new languages and constructs. From Assembly Language to COBOL to Python to Rust, it seems now in retrospect that humans always wanted to communicate with machines in increasing levels of abstractions. Until recently, the ambiguity of natural language -- perhaps its defining hallmark -- has been both the source of its power and at the same time the factor that has limited its use as a vehicle that can convey the exactness required of communicating with a machine. As Brooks noted, one must perform perfectly. This, however, has drastically changed with the arrival of modern generative AI (GenAI). Finally, as Brooks would say, the magic of myth and legend has come true in our time and novice learners can use natural language to instruct computers without the requirements of being ``perfect.''

Although some see the arrival of GenAI as an existential threat to computing education, we argue that this is where we were always headed as a discipline and that it should be embraced. Much like the debates of old about ``objects first'', we argue for a ``prompts first'' approach to programming education. Educators should start courses by teaching students ``prompt engineering'' -- or, more accurately, ``programming in natural language.'' 
Although our call to natural language is motivated by recent GenAI advances, it reflects a perspective argued by Mark Halpern clearly back in 1966 - that \say{natural programming language is one that can be written freely, not just read freely}  \cite{halpern1966foundations}.
As the performance of generative AI continues to increase, the vast majority of programming will focus on problem solving via domain-specific constructs, and not programming-language constructs.

\section{G{\small{en}}AI in Computing Education}

In February 2022, Finnie-Ansley et al. presented a groundbreaking paper at the Australasian Computing Education conference, assessing the capabilities of OpenAI's Codex model for solving introductory programming problems \cite{finnieansley2022robots}. Comparing the performance of Codex to that of students on standard CS1 problems, the model ranked 17th out of 71 students, and was able to produce varied solutions in terms of length, syntax, and algorithmic approach. The authors concluded by expressing their surprise at the model's advanced capabilities, describing the results as ``stunning'' and indicative of an emergent existential threat to traditional methods of teaching and learning introductory programming.

Numerous subsequent papers emerged to present and discuss the many implications of LLMs on computing education \cite{denny2024computing, becker2023programming, prather2023robots}. These have highlighted concerns such as academic integrity, potential over-reliance on AI, and the risk of misleading outputs, while also emphasizing transformative opportunities like generating educational resources and providing instant feedback. The rapid improvement in LLM capabilities, such as the superior performance of GPT-4 over Codex, underscores the potential of these models in education \cite{prather2023robots}.

Educator and student perceptions of LLMs have revealed a mix of excitement and concern. Educators are divided between resisting AI tool usage and embracing them to enhance learning \cite{lau2023banit, sheard2024instructor}. Students appreciate the efficiency of AI tools but worry about over-reliance and the potential for undermining fundamental problem-solving skills \cite{prather2023weird,prather2024widening}. While both groups acknowledge the challenges to learning that GenAI poses, they also recognize the relevance of generative AI tools in computing education and future careers \cite{prather2023robots}.

Leveraging the strengths of generative AI while mitigating its risks will require innovative pedagogical approaches, including teaching students how to work effectively with these models.  The production of learning resources has been well studied already, for example, with Balse et al. \cite{balse2023evaluating} and MacNeil et al. \cite{macneil2023experiences} demonstrating the effectiveness of LLM-generated code explanations in supporting student learning, and Sarsa et al.~\cite{sarsa2022automatic} and Logacheva et al.~\cite{logacheva2024evaluating} documenting the potential for generating novel programming exercises.  More recent efforts have identified the need to help students interact with LLMs, for example, by learning how to create effective prompts \cite{denny2024prompt, nguyen2024how}.

\section{History through the lens of Abstractions}
The history of programming languages is a history of the inventions of abstractions.  
In this section, we discuss a few important historical abstractions, recognizing that it is not an exclusive list.

\subsection{Machine Code}

Machine code enabled the move from electro-mechanical to electronic. It replaced physical patch cables and switches that manipulate hardware with software instructions that manipulate hardware. However, it is important to note that even at the lowest level of abstraction with respect to computer hardware, it is still an abstraction.

\subsection{Low-Level languages}
Assembler provides a symbolic representation layer above machine code.  Each microprocessor supports a custom set of instructions.  Because of the ``nearness'' of the hardware to the language, one can write highly efficient algorithms.  Starting in the 1940s, Kathleen Booth created the assembly language and the design of the assembler for the first ARC computers at Birkbeck college in London~\cite{CCNC2009}.

Now one no longer had to read an opcode number and think ``call'' - one could just write ``call.''  And memory locations could now be named.  Given what programmers wanted to do, having names was better.  It let the programmer get some of the knowledge in the head and put it out into the world, into the program source code.  Now others could better see what the code was supposed to do - some of its hidden intent was made visible, readable to others. 

\subsection{High-Level Languages}

Once we acclimated to luxuries like \emph{comments} and  \emph{macros}, the community did not stay content for long.  Assembler now became known as ``low-level'' programming, ``low'' in the sense of being near the hardware. 
In reflecting on the invention of ``high level languages'' Brooks wrote in 1986, ``Surely the most powerful stroke for software productivity, reliability, and simplicity has been the progressive use of high-level languages for programming. Most observers credit that development with at least a factor of five in productivity, and with concomitant gains in reliability, simplicity, and comprehensibility'' \cite{brooks1986silver}.

The designers of COBOL thought that making the programming language more like natural language of the domain of business would improve it.  While it should be noted that some in the community, such as Dijkstra, fervently opposed COBOL because of its orientation toward the English language, others like Mark Halpern argued that the problem with COBOL was actually that it was not oriented toward English \textit{enough}! Halpern writes of Dijktra's position: ``His diagnosis of the trouble with passive systems is astute, but the cure (insofar as language can offer one) is to make the program easier to write, not harder to read'' \cite{halpern1966foundations}.

It is not difficult to agree with Brooks and Halpern that when a programmer does not have to think about ``decrement a counter, check register and branch-not-zero,'' but instead can think ``loop over these items...,'' that several benefits accrue.

\subsection{Structured Programming}
Structured Programming imposed strict guidelines on high-level languages by removing the go-to statement. Edsger Dijkstra's letter ``Go To Statement Considered Harmful,'' published in CACM in March, 1968 criticized excessive use of ``go to'' and included the admonition ~\cite{dijkstra1968letters}: 
\begin{displayquote}... our intellectual powers are rather geared to master static relations and that our powers to visualize processes evolving in time are relatively poorly developed.  For that reason we should do (as wise programmers aware of our limitations) our utmost to shorten the conceptual gap between the static program and the dynamic process, to make the correspondence between the program (spread out in the text space) and the process (spread out in time) as trivial as possible.
\end{displayquote}
``Shortening the conceptual gap'' evidently does not always involve inventing a new abstraction.  Dijkstra seemed to think it is possible by removing an existing construct, in this case the ``go to'' statement.  

Fast forward now to today's ubiquitous abstraction of the \emph{exception}, an event\begin{displayquote}which occurs during the execution of a program, that disrupts the normal flow of the program's instructions.\cite{javaexception}
\end{displayquote}

It is also referred to as a ``non-local GOTO.''  Dijkstra's critique was taken to heart at the time and then many years later, the languages and conceptual abstractions had matured enough to enable a new way of thinking and implementing a different kind of GOTO statement, a jump ``out, out, out'' and back to the invoking construct.

\subsection{Objects}
Abstract Data Types (ADTs) were an important step in helping a programmer think in terms of higher-level concepts than functions or memory \cite{LiskovZilles1974}.  A \emph{queue} had \emph{attributes} and \emph{methods} and could be implemented several different ways.  Perhaps ADTs helped pave the way to Object Oriented programming \cite{Moon1986}.
Once ``Objects'' were accepted as useful abstractions, educators asked, ``when to teach?'' This discussion became known as ``Objects First,'' some arguing that one should not wait until novices had thoroughly learned a language before introducing object oriented thinking.  

In this light, we phrase our proposal as ``Prompts First,'' arguing that, just like Objects were finally accepted as ``valid,'' one only had to decide when to teach that abstraction.  We claim that large language models are now at that point - having earned a place in programming, it now only remains to ask, ``when to teach?''

%
%

\section{Discussion}

With each increase of the abstraction level over time, programmers were able to focus more on the purpose of the code (what it does) than on the technical details (how it does that). Generative AI continues this process of raising the level of abstraction in current computing constructs. This advancement offers new opportunities for educators to emphasize programming concepts over syntax, overcoming barriers that often demoralize and discourage early programmers~\cite{prather2023robots}. This shift towards prioritizing the purpose of the code can also be more engaging for early students, as it helps them see a clear connection between what they learn and real-world impacts \cite{denny2024explaining}.

However, GenAI also introduces unique challenges. First, it is clear that lower-performing novice programmers can over-rely on it, disrupting their problem solving skills and leaving them with an ``illusion of competence'' ~\cite{prather2024widening}. Second, the mapping between levels of abstraction in Generative AI is non-deterministic. Models can misunderstand natural language inputs and hallucinate incorrect code. 
For these reasons, we must explore and adopt new approaches in both teaching and practical application.

\subsection{Precise Vocabulary}

To address challenges and maximize benefits, a well-defined vocabulary will be essential to convey deeper and more complex concepts. For example, other fields including Mathematics and Music excel in this regard; having a well defined notation to describe concepts. In Mathematics, terms like ``differentiation’' or ``average’' are universally understood within the field and provide a succinct way to describe complex ideas. Similarly, mathematical equations and functions are an unambiguous way to specify an idea. 

Developing and teaching a precise vocabulary will become even more crucial in computer science going forward. As programming in natural language becomes the norm, it will be necessary for students to become more capable of describing the function of their code. Higher level concepts such as mapping, filtering, piping, and exception handling are part of the professional programmer's lexicon because they provide an efficient and unambiguous way to communicate intricate ideas. Educators must focus on instilling this vocabulary in students to equip them for effective communication and problem-solving. In this vein, perhaps Dijkstra was right all along about needing precise language, although perhaps not in the way he thought it would be.

\subsection{Debugging Skills} 

While Generative AI can accelerate coding by generating boilerplate or even complex code snippets, the non-deterministic nature of these models means that errors and unexpected behaviors are inevitable. Therefore, educators must prioritize teaching students debugging techniques. Students should learn to approach debugging with a critical eye, questioning the AI's output and cross-referencing it with their understanding of the problem domain. 

%
%

\section{Conclusion}
A review of programming language history shows that each new language invention raised the level of abstraction.  From dip switches to assembler and on to high-level languages, each new representation enabled us to ignore the ``irrelevant features'' and focus on ``more salient'' ones. Each new representation opened up computer science to a growing audience by removing artificial hurdles.

In \cite{halpern1966foundations}, Halpern's final challenge was this:
\begin{quote}Finally, the possibility of user-guided natural-language programming offers a promise of bridging the man-machine communication gap that is today's greatest obstacle to wider enjoyment of the services of the computer.
\end{quote}

Designing, describing, and processing state and behavior has always been challenging.  But it was made even more challenging than necessary by arbitrary historical representational constraints.  Freed from the previously necessary obsession with \emph{exactly how to do a thing}, today's students can spend more time on salient matters - the constructs of program behavior and the language of the domain. Finally.

\begin{acks}
This research was supported by the Research Council of Finland (Academy Research Fellow grant number 356114).
\end{acks}

\balance

\bibliographystyle{ACM-Reference-Format}
\bibliography{references}

\end{document}